# Fast electron scattering as a tool to study target's structure[*/]

M. Ya. Amusia[1, 2]

[1]Racah Institute of Physics, The Hebrew University, Jerusalem 91904, Israel
[2]A. F. Ioffe Physical-Technical Institute, St.-Petersburg 194021, Russia

**Abstract**

We concentrate on several relatively new aspects of the study of fast electron scattering by atoms and atom-like objects, namely endohedral atoms and fullerenes. We show that the corresponding cross sections, being expressed via so-called Generalized Oscillator Strengths (GOS), give information on the electronic structure of the target and on the role of electron correlations in it. We consider what sort of information became available when analyzing the dependence of GOS upon their multipolarity, transferred momentum $q$ and energy $\omega$.

We demonstrate the role of nondipole corrections in the small-angle fast-electron inelastic scattering. There dipole contribution dominates while non-dipole corrections can be considerably and controllably enhanced as compared to the case of low and medium energy photoionization. We show also that analyses of GOS for discrete level excitations permit to clarify their multipolarity.

The results of calculations of Compton excitation and ionization cross-sections for noble gas atoms are presented.

Attention is given to cooperative effects in inelastic fast electron – atom scattering that results in directed motion of the secondary electrons, a phenomenon that is similar to "drag currents" in photoionization.

We demonstrate how one should derive GOS for endohedral atoms, e.g. $A@C_{60}$ and what is the additional information that can be obtained from corresponding GOS.

Most of discussions are illustrated by the results of concrete calculations.

## 1.  Introductory remarks

Fast electrons interact relatively weak with the target charges. Therefore, the respective inelastic cross-sections, apart of being interesting for different applied areas of science and technology, supply information on the transition probabilities and ionization of the target objects. Thus, they give important knowledge about the electronic wave function of the target in its initial and final states, i.e. before and after interaction with the fast electron.

Since long ago it has been recognized that the fast charge particle (including electrons) inelastic scattering cross-section can be presented as a product of two factors solely dependent upon the properties of the projectile and target, respectively [2]. All features of the target are collected in the so-called Generalized Oscillator Strengths (GOS). Their definition and main

---

[*]/It is my pleasure indeed to present this paper, in fact a short review, to a volume of Proceedings of a Symposium dedicated to the 70[th] birthday of Prof. Dr. Russell A. Bonham. His great contribution to the field of theoretical and experimental study of the excitation and ionization of atoms and molecules using electron scattering and coincidence techniques is widely acknowledged. His papers in the field were published already as earlier as more than forty years [1]

properties are discussed in monographs (e.g. [3]) and described at length in review articles (e.g. [4])

The use of fast electrons as a tool to study the targets internal structure has definite advantages as compared to photoionization that is also powerful in investigating the structure of the target. The main advantage is the ability of fast electrons to excite not only dipole transitions, but also transitions of other multipolarity. It is also essential that the GOS are, unlike photoionization cross-sections, dependent not only upon the transferred to the atom energy $\omega$ but also transferred momentum $q$. As a result the GOS are able to check the targets wave functions at different distances.

It is essential to have in mind that the fast electron inelastic scattering cross-sections are much bigger than that of photoionization. On the other hand, the disadvantage of inelastic electron scattering as compared to photoionization is the relatively big role played by interaction between the projectile and target before and after the target ionization or excitation takes place.

Although known since long ago and being a relatively simple method of receiving information on targets structure, even called "synchrotron for pure men", the inelastic fast electron scattering still has a lot of resources. The ability of this approach is far from being exhausted. A lot was done recently in this domain, including problems of small angle scattering, minima in GOS etc, which is illustrated by the prominent number of recent publications (see e.g. [5-11]).

The content of this brief review will be some of the recent results obtained in this direction by my colleagues and me, and outlining some nearest perspectives in this area of research.

By studying GOS one can obtain information on electron correlations of different multipolarity, which is almost impossible in photoionization studies. Indeed, we will present results demonstrating for noble gas atoms the big difference [12] between one-electron GOS obtained in Hartree-Fock (HF) approximation, and that taking into account the deviation from the one-electron picture in the frame of the Random Phase Approximation with Exchange (RPAE) [13, 14]. The correlations proved to be important for dipole and at least for quadrupole channels of the target ionization in a big domain of $q$ [12].

It was demonstrated recently, that it is great interest to study photoionization in atoms with semi-filled electron subshells, where strong dependence of the cross-sections upon term of the residual ion was found [15]. We suggest that the same problem could be of interest in GOS studies. Of special interest is the case of GOS for 3$d$ electrons in Xe, Cs and Ba, where the role of $3d_{3/2}$ level upon $3d_{5/2}$ was clarified and proved to be very important [16], similar to the case of photoionization [17]. Note that in calculations the $3d_{5/2}$ and $3d_{3/2}$ levels were treated as semi-filled subshells.

It appeared that by investigating GOS of discrete levels we can determine their multipolarity or disclose the presence of e.g. two closely located discrete levels, otherwise almost indistinguishable [18, 19]. As it was demonstrated in [19], the experimentally observed 3p-4p level is mainly monopole, while the experimentalists claimed that it is quadrupole [20]. It appeared that the RPAE correlations affect considerably the position and magnitude of the minima in GOS of discrete levels [21].

Note that at small $q$ GOS tend to ordinary dipole oscillator strengths (OSO) at any projectile energy [22]. With growth of $q$ non-dipole corrections become increasingly important. Although they can be detected in the angular distributions of photoelectrons even at not too high photon frequencies $\omega$ [23], the relative role of these corrections is much smaller there than it can be in GOS. In the latter case varying the transferred momentum can controllably change the non-dipole terms contribution.

Similar to the case of photoionization, inelastic electron scattering leads to creation of the so-called "drag current" [24] – a coherent cooperative movement of ionized electrons, the estimation of which will be performed in this paper.

Compton scattering and excitation cross-sections can be expressed via respective GOS. We will present results obtained for Compton scattering on noble gases [25] and on discrete levels excitation [26].

We will present semi- qualitative analyses of GOS for endohedral atoms, e.g. $A @ C_{60}$, at first, the manifestation of so-called confinement resonances, which were discussed recently in photoionization [27]. We will analyze also another effect of $C_{60}$, namely the influence of its dynamic polarization [28] upon the dependence of GOS of the atom $A$ located inside the fulleren shell upon frequency and angular momentum.

Atomic System of units, with electron mass $m$, charge $e$, and Planck constant $\hbar$ equal to 1, $m = e = \hbar = 1$ is used in this paper.

## 2. Main formulas

Differential in transferred to the target atom energy $\omega$ and incoming fast particle scattering angle $d\Omega$, the inelastic scattering cross-section $d^2\sigma_{if}(\omega,q)/d\omega d\Omega$ that is accompanied by the target's transition from the initial state $i$ to the final $f$ is given by the following relation [2,3]:

$$\frac{d^2\sigma_{if}}{d\omega d\Omega} = \frac{4p'}{q^2 p\omega} G_{fi}(\omega, q). \tag{1}$$

Here $p, p'$ are the momenta of incoming and outgoing projectile, $q = |\mathbf{p} - \mathbf{p}'|$, and the GOS $G_{fi}(\omega, q)$ is defined by the following formula:[2, 3]

$$G_{fi}(\omega, q) = \frac{2\omega}{q^2} \left| \sum_{j=1}^{N} \int \Psi_f^*(\vec{r}_1,...,\vec{r}_N) e^{i\vec{q}\vec{r}_j} \Psi_i(\vec{r}_1,...,\vec{r}_N) d\vec{r}_1...d\vec{r}_N \right|^2, \tag{2}$$

where $N$ is the number of atomic electrons and $\Psi_{i,f}$ are the atomic wave functions in the initial and final states with energies $E_i$ and $E_f$, respectively; $\omega = E_f - E_i$. Note that the final state can belong to both discrete and continuous spectrum. Because the projectile is assumed to be fast, its wave functions are plane waves and its mass $M$ enters GOS indirectly, namely via the energy and momentum conservation law:

$$\frac{p^2}{2M} - \frac{p'^2}{2M} \equiv \frac{p^2}{2M} - \frac{(\mathbf{p} - \mathbf{q})^2}{2M} = \omega. \tag{3}$$

Here $p$ is the momentum of the projectile in the initial state.

It follows from GOS definition (2) that when $q = 0$ GOS coincides with the ordinary oscillator strength (OOS) of optical transitions or is simply proportional to the photoionization cross section (see, for example [3]), depending upon whether the final state is a discrete excitation or belongs to the continuous spectrum. The energy $\omega$ enters GOS either via a factor in Eq. (2) or indirectly, via the energy $E_f$.

Let us expend the exponent $\exp(i\vec{q}\vec{r})$ in (1) into the following series

$$e^{i\vec{q}\vec{r}} = \sum_{L=0}^{\infty} i^L (2L+1) j_L(qr) P_L(\cos\vartheta). \tag{4}$$

Then in one-electron Hartree-Fock (HF) approximation Eq. (2) with the help of (4) simplifies considerably, reducing to the following relation:

$$g_{fi}^L(\omega, q) = \frac{2\omega}{q^2} \left| \int \phi_f^*(\vec{r}) j_L(qr) P_L(\cos\vartheta) \phi_i(\vec{r}) d\vec{r} \right|^2 \equiv \frac{2\omega}{q^2} \left| \langle f | j_L(qr) | i \rangle \right|^2, \tag{5}$$

where $\phi_{f,i}(\vec{r}) = R_{\varepsilon(n_f),i}(r) Y_{l_{f,i}, m_{f,i}}(\theta_r, \varphi_r) \chi_{s_{f,i}}$ are the HF one-electron wave functions with their radial, angular and spin parts, respectively, $j_L(qr)$ is the spherical Bessel function, $P_L(\cos\vartheta)$ is the Legendre polynomial and $\cos\vartheta = \vec{q}\vec{r}/qr$. The excitation energy of the $i \to f$ transition is denoted as $\omega_{fi}$. The principal quantum number, the angular momentum, its projection and spin quantum numbers of the initial $i$ and final $f$ states are denoted by $n_{f,i}$, $l_{fi}$, $m_{f,i}$ and $s_{f,i}$, respectively. Final state continuous spectrum wave functions are determined instead of $n_f$ by one-electron energy $\varepsilon$.

The multi-electron correlations we will take into account in the frame of well-known Random Phase Approximation with Exchange - RPAE. In the language of diagrams the inelastic scattering of fast electrons in the RPAE frame can be presented in the following way [13]:

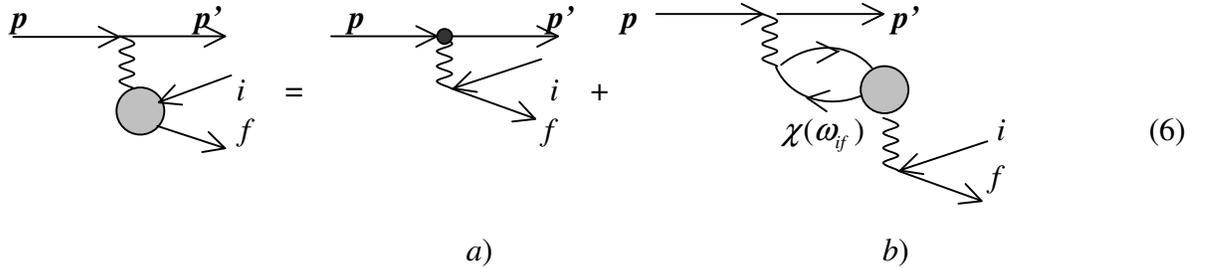

$$\hspace{10em} a) \hspace{15em} b) \hspace{10em} (6)$$

The dashed line, the line with an arrow to the right (left) and the wavy line represent the incoming photon, electron, vacancy and the Coulomb interelectron interaction, respectively. The gray circle stands for the effective interaction between the atom and the virtual longitudinal photon emitted by the projectile.

Analytically, the RPAE equations look as follows [13]:

$$\langle f | A_L(q, \omega_{fi}) | i \rangle = \langle f | j_L(qr) | i \rangle + \left( \sum_{n' \leq F, k' > F} - \sum_{n' > F, k' \leq F} \right) \frac{\langle k' | A_L(q, \omega_{fi}) | n' \rangle \langle n'f | U | k'i \rangle_L}{\omega_{fi} - \varepsilon_{k'} + \varepsilon_{n'} + i\eta(1 - 2n_{k'})}. \tag{7}$$

Here $\leq F (> F)$ denotes occupied (vacant) HF states, $\varepsilon_n$ are the one-electron HF energies, $\eta \to 0$ and $n_k = 1(0)$ for $k \leq F (> F)$; $\langle nf | U | ki \rangle_L = \langle nf | V | ki \rangle_L - \langle nf | V | ik \rangle_L$ is the $L$ component of the matrix elements of the Coulomb interelectron interaction $V$ (see e.g. [14]). It is seen that the equation for each total angular momentum of an excitation $L$ is separate. The procedure of solving this equation is considered in details in [13, 14]. System of equations (7) is solved numerically, as is described in [14].

A relation similar to (5) determines the GOS in RPAE $G_{fi}^L (q, \omega_{fi})$:

$$G_{fi}^L (q, \omega_{fi}) = \frac{2\omega_{fi}}{q^2} \left| \langle f | A_L (q, \omega_{fi}) | i \rangle \right|^2, \qquad (8)$$

where $\langle f |$ and $| i \rangle$ are the final and initial HF states, respectively.

Symbolically, the RPAE equations can be presented as is done in [13, 14]

$$\hat{T} = \hat{t} + \hat{T} \hat{\chi} U, \qquad (9)$$

where $U$ is the combination of Coulomb direct and exchange interelectron interaction, and $\hat{\chi}$ is the term, describing virtual electron – vacancy excitation [see $\chi(\omega_{if})$ in (6)], which can be presented symbolically as follows:

$$\hat{\chi} = \hat{1}/(\omega - \omega' + i\eta) - \hat{1}/(\omega + \omega') \qquad (10)$$

with $\eta \to +0$ and $\omega$ being the energy, transferred to the target atom from the projectile, while $\omega'$ is the energy of any atomic discrete or continuous spectrum electron – vacancy excitation.

The equations (6) and (7) determine the RPAE values for the GOS of continuous spectrum and discrete excitations. Some time ago calculations of GOS for continuous spectrum were performed for the following subshells of noble gas atoms: $2s^2$ and $2p^6$ in Ne, $3s^2$ and $3p^6$ in Ar, $3d^{10}$, $4s^2$ and $4p^6$ in Kr, $4d^{10}$, $5s^2$ and $5p^6$ in Xe [12]. The results were obtained in one-electron HF approximations and with account of multi-electron correlations in the frame of RPAE. The transferred linear momentum $q$ varied from zero to 2 atomic units and the transferred energy $\omega$ varies from ionization threshold to 5-8Ry. We took into account four values of the transferred angular momentum $L = 0 \div 3$. It appeared that the role of RPAE corrections is the biggest in the dipole channel, $L=1$, but also quite noticeable in the quadrupole channel, $L=2$, also.

### 3. Intradoublet correlations

It was demonstrated recently that due to interaction between electrons belonging to two components of the spin-doublet, $3d_{5/2}$ and $3d_{3/2}$ in Xe, Cs and Ba, the partial photoionization cross-section $\sigma_{3d5/2}$ acquires a prominent maximum that was called intradoublet resonance [17]. This behavior is reflected also in other characteristics of the photoionization process.

Of interest is to investigate how the intra-doublet interaction is reflected in GOG, namely in its dipole component as a function of $q$ and learn how this interaction is manifested in transitions of other multipolarity, namely monopole, quadrupole and octupole. This rather complicated calculation can be considerably simplified if one treats electrons belonging to $3d_{5/2}$ and $3d_{3/2}$ levels as, approximately, two closed subshells of equal number of spin "up" and "down" electrons.

Corresponding generalization of (7) to a system of two kinds of electrons was developed in [29] and adjusted to the case on $3d$ spin doublet in [17]. Another effect beyond RPAE that is essential for deep subshells like $3d$ in Xe, Cs and Ba is rearrangement, that is the modification of the outer electrons' state while the inner electron leaves the atom. All this leads to the approximation, the name of which has been abbreviated as SP GRPAE that means Spin Polarized Generalized RPAE.

Using just this approach, the GOS for dipole, monopole and quadrupole transitions were calculated in [16] for both, $3d_{5/2}$ and $3d_{3/2}$, levels in Xe, Cs and Ba for energies $\omega$ above the $3d$-threshold by 60-80 eV and in the momentum $q$ range $0.1 < q < 4$ a.u. (atomic units). Particularly strong modifications under the action of the intra-doublet interaction were found in the dipole transition that even increase with $q$ growths. In general, the GOS of $3d_{5/2}$ electrons are strongly affected, while $3d_{3/2}$ GOS are almost not affected by electron correlations. To illustrate these statements, in Fig. 1 we present the results for GOS of Cs $3d$ electrons[16]. It is seen at the same time that the height of the maxima are decreasing with $q$ growth.

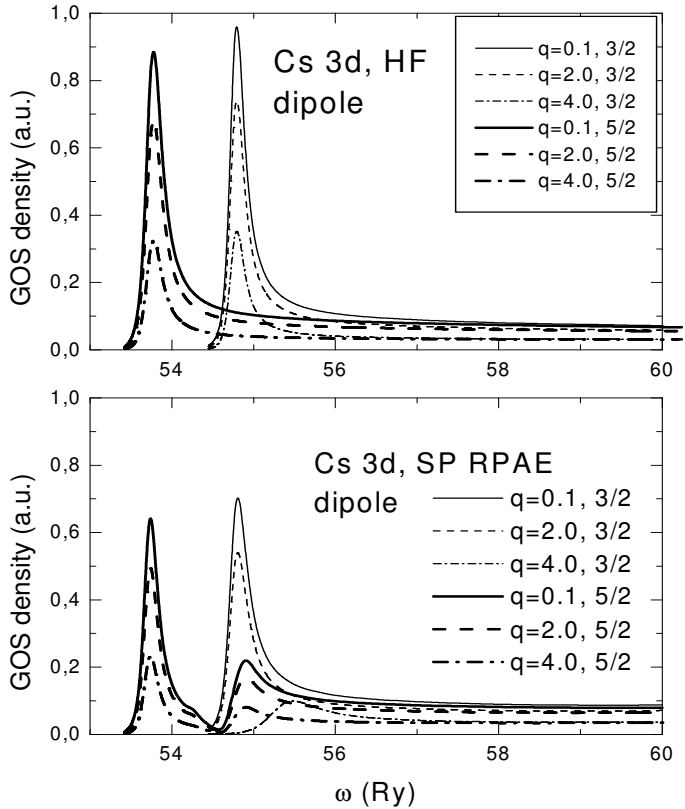

Fig. 1. Dipole generalized oscillator strengths for Cs $3d$ subshell: (a) HF approximation and (b) SP RPAE results.

Fig. 2 illustrates the situation with the monopole GOS [16]. It is seen, that while they are strongly affected by intradoublet interaction, they are by an order of magnitude smaller than the dipole GOS. As to the GOS of quadrupole transition, they proved to be considerably less sensitive to the intradoublet interaction than dipole and monopole GOS. Their magnitude in the considered in this section cases is extremely small: three orders of magnitude less than the monopole GOS.

Note, that experimental investigation of considered above GOS is of interest and significance as a test of our ability to understan the evolution of the inradoublet resonance with growth of the momentum $q$ transferred to the atom in the inelastic collision process. It requires, however, rather complicated experiment, in which the inelastic scattered and eliminated from atom electrons would be detected in coincidence.

## 4. Identification of discrete levels

An interesting application of GOS studies is the detection of the angular momentum of discrete transitions. Relatively recently the GOS of the lowest non-dipole transition in Ar, $3p \rightarrow 4p$, was measured [20] and on the ground of obtained data has been identified as quadrupole. It was soon demonstrated that the experimental accuracy achieved in [20] is not enough to separate quadrupole and monopole $3p \rightarrow 4p$ levels, and in fact only GOS of their mixture was measured [18].

Calculations were performed by solving (7) and (8) numerically.

There is a specific feature that has to be taken into account when (7) is applied to discrete transitions. Namely, the denominator in the second term from the right hand side of (7) diverges at $\omega_{fi} = \varepsilon_f - \varepsilon_i$. To overcome this divergence, one has to isolate this term from the summation and take it into account analytically [14]. This step leads to alteration of the discrete excitation energy, in our case $\omega_{fi}$, from being simply equal to $\omega_{if} = \varepsilon_f - \varepsilon_i$. The real excitation energy $\tilde{\omega}_{if}$ is given by the sum of $\omega_{if}$ and the matrix element of effective "electron *f*-vacancy *i*" interaction $<i, f | \Gamma(\omega_{fi}) | i, f>$, the latter in general being essentially different from pure Coulomb interaction *U*. This matrix element was calculated according to procedure described in [14].

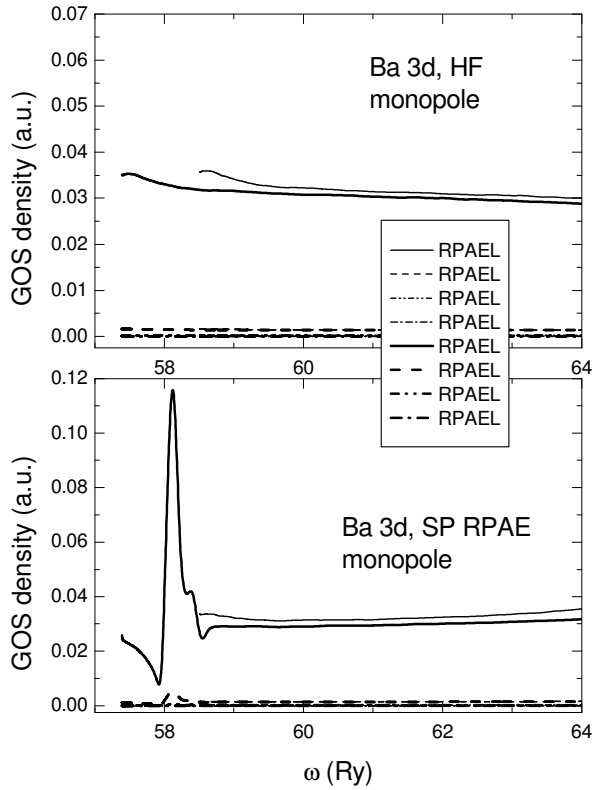

Fig. 2. Monopole generalized oscillator strengths for Ba *3d* subshell: (a) HF approximation and (b) SP RPAE results.

After eliminating the divergent term from the sums in (7), the equation is solved leading to matrix elements $<i | \tilde{A}(q,\omega) | f >$. As it was demonstrated in [14], due to $\omega$-dependence of $\Gamma(\omega)$, the expression for discrete excitation GOS are given by expression similar to (8), with an additional factor $Z_{if}$:

$$G_{fi}^L(q,\tilde{\omega}_{fi}) = Z_{if}\frac{2\tilde{\omega}_{fi}}{q^2}\left|\left\langle f\left|\tilde{A}_L(q,\tilde{\omega}_{fi})\right|i\right\rangle\right|^2, \tag{11}$$

where the renormalization factor $\tilde{Z}_{if}$ is given by the following relation [14]:

$$Z_{if} = \left[1 - \partial <i,f|\Gamma(\omega)|i,f>/\partial\omega|_{\omega=\omega_{if}}\right]^{-1} \tag{12}$$

In our calculations [18, 19], however, $\tilde{Z}_{if}$ proved to be close to one.

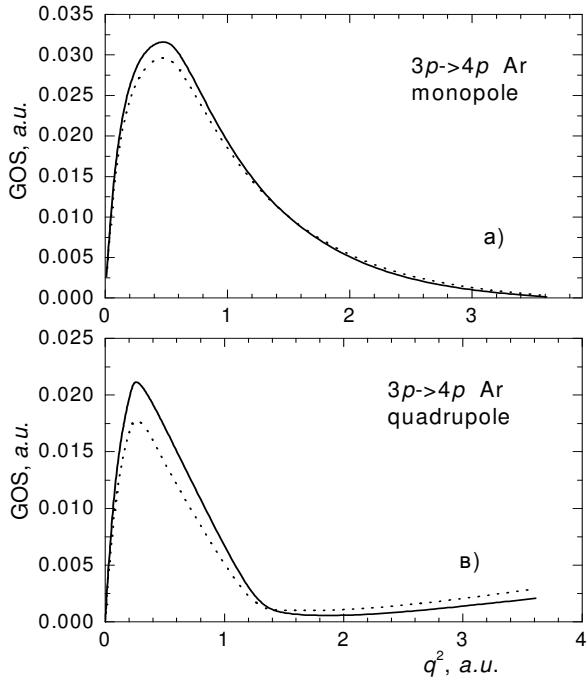

Fig. 3. HF (dots) and RPAE (heavy line) GOS of 3p-4p transitions. (a) – monopole, (b) – quadrupole [18]

The results of calculations are presented in Fig. 3 and 4. It is seen from Fig.3 that the monopole contribution is considerably bigger than the quadrupole one.

Fig. 4 demonstrates that only by taking into account the contribution of both monopole and quadrupole transitions reasonable agreement with experiment can be achieved.

Note, that the GOS behavior studied in [19] for lowest excitations of all noble gases $np \rightarrow (n+1)p, (n+2)p$ is similar to that presented in Fig. 3. Namely, the difference between quadrupole and monopole GOS is noticeable: the monopole is bigger, at the main maximum by a factor of 1.5-1.8. As to the quadrupole GOS, it has a second maximum at relatively high $q^2$.

Since the quadrupole levels are decaying faster than the monopole ones, namely by emitting one quadrupole photon instead of two dipole photons, the excited atoms with remain in their monopole state. It seems that by measuring the quadrupole photon yield one can reliably distinguish the quadrupole excitations from the monopole.

Of particular interest is the consideration of dipole discrete excitation GOS, e.g. 3p-4s in Ar, where long ago a minimum was predicted [1] and much later found in experiment [30]. Note that quite recently it was demonstrated [31] that the electron correlations are not important while exchange is important in location of the position and value of the GOS minimum in Ar 3p-4s excitation.

The GOS of outer shell discrete dipole excitations, e.g. 3p-4d in Ar, are of interest since are very close in energy to octupole excitations with the same configuration. While at very small $q^2$ the dipole contribution absolutely dominates, the quadrupole GOS increases very fast, so that already at $q^2 \geq 2$ it is much bigger than the dipole [32]. This feature is typical for all noble gas atoms [32]. The high probability of octupole excitations at not too big $q^2$ opens the possibility of creating a gas volume with a large number of atoms in octupole-excited states.

## 5. Small *q* non-dipole corrections

Investigation of non-dipole transitions in angular distributions of photoelectrons permitted experimental investigation of atomic quadrupole continuous transition matrix elements [33]. The investigations were inspired by the following expression for the angular distribution of photoelectrons from the *nl* subshell [34, 35, 36, 24]:

$$\frac{d\sigma_{nl}^{\gamma}(\omega)}{d\Omega} = \frac{\sigma_{nl}^{\gamma}(\omega)}{4\pi}[1 - \frac{\beta_{nl}(\omega)}{2}P_2(\cos\theta) + \kappa\gamma_{nl}(\omega)P_1(\cos\theta) + \kappa\eta_{nl}(\omega)P_3(\cos\theta)], \qquad (13)$$

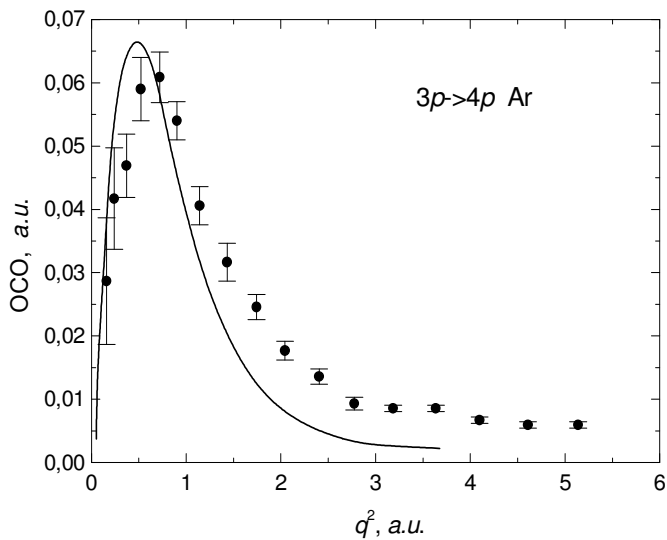

Fig. 4. Experimental (dots) [20] and calculated [18] GOS (sum of monopole and quadrupole contributions) of 3*p*-4*p* transition in Ar.

where $\sigma_{nl}^{\gamma}(\omega)$ is the *nl* subshell photoionization cross-section, $\kappa = \omega/c$, *c* is the speed of light, $P_l(\cos\theta)$ are the respective $l = 1, 2, 3$ Legendre polynomials, $\theta$ is the angle between photoelectron and incoming photon momenta; the parameters $\beta_{nl}(\omega)$, $\gamma_{nl}(\omega)$ and $\eta_{nl}(\omega)$ are expressed via dipole $nl \to \varepsilon l \pm 1$ and quadrupole $nl \to \varepsilon l, l \pm 2$ transitions matrix elements and $l, l \pm 1, l \pm 2$ photoelectron scattering phases [24].

The prominent defect of using equation (13) as a tool to study quadrupole continuous spectrum transition originates from the smallness of the parameter $\kappa$ in the non-relativistic domain, in particular close to thresholds of the outer atomic shells, where $\kappa \approx 1/c \ll 1$. It is essential also, that equation (13) does not include any information about monopole excitations.

Both these minuses could be eliminated if to consider small *q* inelastic fast electron scattering by atoms. To illustrate this, let us concentrate for simplicity on ionization of an s-subshell only. For this case the following formula can be obtained for the differential in angle of the ionized atomic electron cross-section, if one limits consideration with monopole, dipole and quadrupole transitions only[*/]:

---

[*/] The following expressions were derived and analyses performed together with A. S. Baltenkov.

$$\frac{d^2\sigma}{d\Omega_k dq} = \frac{1}{4\pi}\frac{d\sigma}{dq}\sum_{j=0}^{3} B_j P_j(\vartheta), \tag{14}$$

where $\sigma \equiv d^2\sigma_{if}(\omega,q)/d\omega d\Omega$ from (1) and

$$B_0 = 1, \quad B_1 = -\frac{6Q_0 Q_1 \cos(\delta_0 - \delta_1) + 12 Q_1 Q_2 \cos(\delta_1 - \delta_2)}{Q_0^2 + 3Q_1^2 + 5Q_2^2},$$

$$B_2 = \frac{2}{7}\frac{1}{Q_0^2 + 3Q_1^2 + 5Q_2^2}[21Q_1^2 + 25Q_2^2 + 35 Q_0 Q_2 \cos(\delta_0 - \delta_2)]$$

$$B_3 = -\frac{18 Q_1 Q_2 \cos(\delta_1 - \delta_2)}{Q_0^2 + 3Q_1^2 + 5Q_2^2}, \quad B_4 = \frac{90}{7}\frac{Q_2^2}{Q_0^2 + 3Q_1^2 + 5Q_2^2}. \tag{15}$$

Here the following notations are introduced $Q_L \equiv Q_{L,n\varepsilon}(\varepsilon,q) = \int R_{n0}(r) j_L(qr) R_{\varepsilon L}(r) r^2 dr$ and $P_i(\vartheta)$ are the Legandre polynomials of the cosine of the angle between vectors of the transferred to the atom momentum **q** and outgoing electron momentum **k**, $k = \sqrt{2\varepsilon}$; $\delta_l$ are the knocked-out electron's scattering phases. We see that $Q_L$ is just the first matrix element in the right hand side of (7).

For small $q$ dominates the dipole term $Q_1 \sim q$, while $Q_0, Q_2 \sim q^2$, with in general different coefficients of proportionality. Note that in electron scattering small $q$ small $q < 1$ that can be much bigger than $1/c$. At $1/c \ll q \ll 1$ only terms $B_0, B_1, B_2, B_3$ "survive", becoming except for $B_0 = 1$ considerably simpler than in (15). To take the RPAE correlations into account one has to substitute $Q_L$ with corresponding solutions of (7) $\tilde{Q}_L = |\tilde{Q}_L| \exp(i\Delta_L)$, where $\Delta_L \equiv \Delta_L(q,\varepsilon)$. Then instead of (15), one has:

$$B_1 = -6\frac{|Q_0|\cos(\tilde{\delta}_0 - \tilde{\delta}_1) + 2|Q_2|\cos(\tilde{\delta}_1 - \tilde{\delta}_2)}{|Q_1|}, \quad B_2 = 2, \quad B_3 = -6\frac{|Q_2|}{|Q_1|}\cos(\tilde{\delta}_1 - \tilde{\delta}_2), \tag{16}$$

where $\tilde{\delta}_L \equiv \delta_L + \Delta_L$.

It is seen that inelastic electron scattering is able to supply information on the monopole and quadrupole parameters simultaneously. It is important that by changing $q$ one can control the contribution of the non-dipole terms and thus enhance considerably the ability to measure them. However, in order to do this coincidence experiment is required, in which simultaneously two final-state electrons, the fast inelastic scattered and the relatively slow removed from the target, would be detected.

The relative contribution of non-dipole terms can be essentially enhanced due to the presence of monopole and quadrupole autoionization resonances, and suppressed by the dipole resonances.

With growth of $q$ up to $q \sim 1$ the simple formula (16) is no more valid and the general expressions (15) have to be used.

## 6. "Dragging" of secondary electrons

Similar to the case of photoionization [34, 24], where the term proportional to $P_1(\cos\vartheta)$ in (13) leads to forward-backward asymmetry, is the case of fast-electron small-angle scattering. In the case of photoionization this asymmetry results in so-called "drag currents" presenting directed motion of photoelectrons in a gas volume irradiated by a beam of photons. The current $J(\omega)$ is determined by the following formula [34, 24]:

$$J_{nl}(\omega) = -WS\kappa\gamma_{nl}(\omega)\frac{1}{3}\frac{\sigma_{nl}^\gamma(\omega)}{\sigma_{eA}(\varepsilon_n)}. \qquad (17)$$

Here $W$ is the photon beam's intensity and $S$ is its cross-section, $\sigma_{nl}^\gamma(\omega)$ is the photoelectrons' scattering cross-section by target atoms, $\varepsilon_{nl} = \omega - I_{nl}$, and $I_{nl}$ is the $nl$ subshell ionization potential. Other notations are the same as in (13). A crude estimation gives for synchrotrons $WS \approx 10^{12} s^{-1}$, leading to currents of about $10^{-11} \div 10^{12} A$.

Since $\gamma_{nl}(\omega)$ as a function of $\omega$ can change sign and is strongly affected by the RPAE correlations, the same is characteristic for the current that is according to (17) independent upon the target gas density. The current can be essentially amplified at autoionization dipole and quadrupole resonances. The so-called Ramsauer minima in $\sigma_{eA}(\varepsilon_n)$ also amplify the current, since they determine the resistance of the gas to the free flow of photoelectrons.

Slow electrons knocked-out off the target in the process of fast–electron inelastic scattering, could also form a drag current. For a given, but small $q$, the contribution to this current from ionization of $nl$ subshell is determined by an equation similar to (17):

$$J_{nl}(\omega, q) = -WSB_1\frac{1}{3}\frac{d\sigma_{nl}(\omega, q)/dq}{\sigma_{eA}(\varepsilon_{nl})}. \qquad (18)$$

It is essential to have in mind that according to (15) the parameter $B_1$ for $nl$ subshell depends upon $q$ and $\omega$: $B_1 \equiv B_{1;nl,\varepsilon_{nl}L\pm 1}(\omega, q)$. More suitable for experimental detection in electron-atom scattering would be currents summed over all subshells and integrated over $q$ and $\omega$. The flux $W$ in electron scattering case is much bigger than in photoionization, just as $B_1 \gg \kappa\gamma_{nl}$ and $\sigma_{nl}(\omega, q) \gg \sigma_{nl}^\gamma(\omega)$. As a result, rather big currents, much more than the above mentioned $10^{-11}A$, could be expected. However, in spite of obvious interest, this area is not yet developed neither theoretically nor experimentally.

## 7. Compton scattering

Compton scattering is a process of inelastic photon scattering. Its cross-section for non-relativistic electrons is determined by the second-order in $1/c$ photon-electron interaction operator $\vec{A}^2(\vec{r})/2c^2$ [37]. Since for the external electromagnetic field one can choose the vector potential as $\vec{A}(\vec{r}) = \vec{e}\exp(i\vec{\kappa}\vec{r})$, that leads to the following expression for the differential in the photon emission angle $\Omega^\gamma$ Compton scattering cross-section $d\sigma_{if}^C(\omega_{fi})/d\Omega^\gamma$:

$$\frac{d\sigma^C_{if}(\omega_{fi})}{d\Omega^\gamma} = \left(\frac{d\sigma^C}{d\Omega^\gamma}\right)_{cl} \frac{E-\omega_{fi}}{E} \frac{q^2}{2\omega_{fi}} \sum_f \int G_{fi}(\omega_{fi},q) \equiv \left(\frac{d\sigma^C}{d\Omega^\gamma}\right)_{cl} \xi(\omega_{fi},q) \qquad (19)$$

Here $(d\sigma^C/d\Omega_\gamma)_{cl}$ is the classical Thompson scattering cross-section of a photon upon an electron [37], $q = |\vec{k}-\vec{k}'|$, $k'$ is the emitted photon momentum, $k'=(E-\omega_{fi})$. Comparing (1) and (19), we see that the results for fast electron inelastic scattering and Compton scattering are interconnected.

Compton scattering is of particular interest at high photon energies $E$, where it eventually becomes bigger than the photoionization cross-section. For He it happens already at $6 keV$. This value rapidly increases with atomic weight. Therefore one has the following limitation $\omega_{fi}/E \ll 1$, since atomic structure is essential for excitations of the ionization potential $I$ order, $\omega_{fi} \sim I$. Neglecting terms with $\omega_{fi}/E \ll 1$ in powers higher than one, the following relation can be obtained:

$$q = \frac{2E}{c}\left(1-\frac{\omega_{fi}}{2E}\right)^{1/2} \sin\vartheta^\gamma \approx \frac{2E}{c}\sin\vartheta^\gamma, \qquad (20)$$

where $\vartheta^\gamma = \theta^\gamma/2$ and $\theta^\gamma$ is the scattering angle of the outgoing photon. Thus, for a given value $E$ the angle $\theta^\gamma$ determines the momentum $q$ and vice versa.

Recently, numerical calculations were performed for Compton scattering cross-sections [25, 26] of ionization and discrete excitations of several lowest levels. Noble gas atoms Ne, Ar, Kr and Xe were considered. Contributions were taken into account of transitions with transferred angular momentum in the range $L = 0 \div 3$ and $q \leq 8$ at. un. The results are exemplified in Fig. 5 and Fig. 6.

Fig. 5 presents the results for Compton ionization of Kr 3$d$ electrons. It is seen that the cross-section is dominated by a powerful maximum that is preceded by a minimum. It appeared that with the growth of $\omega$ bigger and bigger $q$ values became increasingly important. In order to calculate the total Compton cross-sections one needs to know the GOS values at high q. In the process of calculations it became clear that for such cross-sections we need to have GOS that correspond to higher than $L = 3$ transferred angular momenta.

Fig. 6 depicts the Compton cross-section of excitation of Kr 4p-5p two very close located levels, namely monopole and quadrupole. Monopole cross-section has one big maximum, while the quadrupole has two. Experimentally, they are almost inseparable at high $E$, and therefore the sum of monopole and quadrupole terms is also presented in the figure.

8. **GOS of endohedral atoms**

Since recently a lot of attention is given to studies, yet at this moment only theoretical, of photoionization of so-called endohedral atom. This is an atom A located in the empty space of a fullerene $F$, that is a highly symmetric structure of carbon atoms C located at a surface. Endohedral atom is denoted as $A@F$. Best known is the fulleren $C_{60}$. In essence, the system

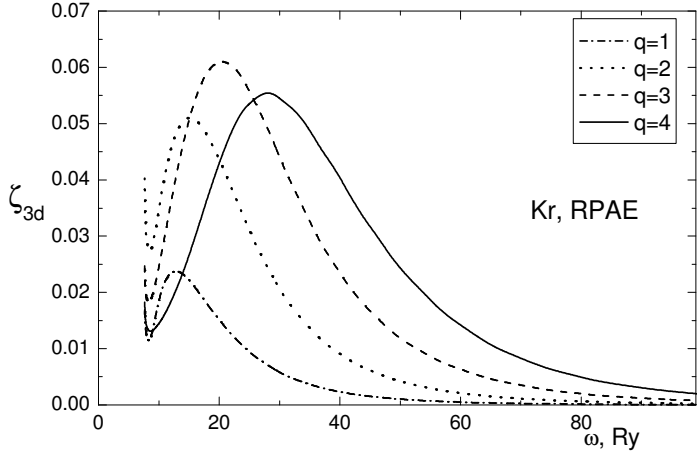

Fig. 5. Parameter $\xi_{3d}(\omega, q)$ from (19) for 3d electrons of Kr as a function of two variables.

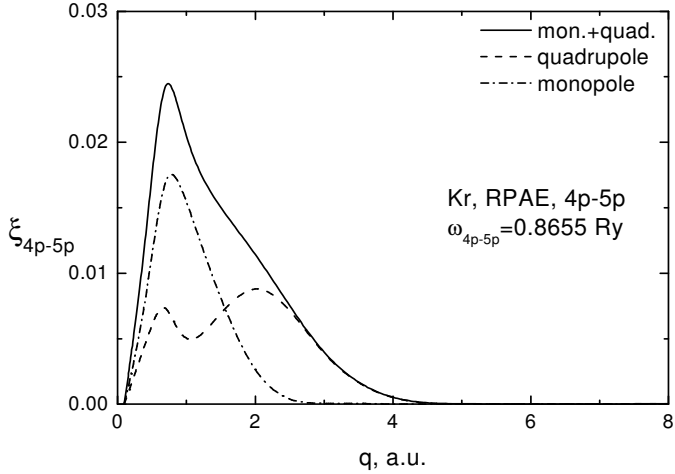

Fig. 6. Parameter $\xi_{4p-5p}(\omega_{4p-5p}, q)$ from (19) for 3d electrons of Kr as a function of momentum $q$.

$A@F$ can be considered as an extended artificial atom, which is formed by an extra really multi-electron shell, added to A.

Just as in an ordinary atom, the inter-shell and intra-shell correlations [13] are essential and can be treated similarly to ordinary subshells of many-electron atoms.

The fulleren affects and is affected by the atom A, modifying its properties and cross-sections. Most prominent are the modification of the inner atom A ionization cross-section due to two factors.

The first factor is the reflection of the ionized from atom A electron by the fullerenes shell: This factor manifests itself in interference patterns in the ionization cross-section as a function of the energy $\varepsilon$ of the electron, ionized from $F$. The second effect is the modification of the interaction between the projectile and atom A due to virtual or real excitations of the fulleren $F$ electron shell. The detailed description of both effects in connection to photoionization one can find in [38]. Therefore here we will limit ourselves with repeating only the main points of the methods employed in [38] emphasizing essential differences between photoionization of and the fast electron inelastic scattering on the $A@F$ atom.

The atom's A radius is much smaller than the radius $R$ of the fullerene shell. Its thickness $\Delta$ is also small, as compared to $R$, $\Delta \ll R$. Therefore, for slow ionized electrons the real complex fulleren potential for a spherically-symmetric $F$ such as e.g. $C_{60}$ can be substituted by a simple pseudo-potential:

$$V(r) = -V_0 \delta(r - R). \tag{21}$$

The radius $R$ is known from experiment, while $V_0$ can be determined to describe electron affinity of $C_{60}^{-}$ in accord with observed data [39].

The effect of (21) can be taken into account analytical, by expressing the partial wave with angular momentum $l$ via the so-called regular $u_{kl}(r)$ and irregular $v_{kl}(r)$ solutions of the atomic Hartree – Fock equation [14] for an ionized electron with linear momentum $k = \sqrt{2\varepsilon}$

Inclusion of (21) leads to a factor $F_{l'}(k)$ in the photoionization amplitude that depends only upon the photoelectron's linear $k$ and angular $l'$ moments [39, 27]:

$$F_l(k) = \cos \delta_l(k) [1 - \tan \delta_l(k) v_{kl}(R) / u_{kl}(R)], \tag{22}$$

where $\delta_l(k)$ is the ionized electron's elastic scattering phase shift that can be expressed by the following relation:

$$\tan \delta_l(k) = u_{kl}^2(R) / [u_{kl}(R) v_{kl}(R) - k / 2V_0]. \tag{23}$$

Using (23), the following relation for the partial photoionization cross-sections of the endohedral atom $\sigma_{kl,l'}^{A@F}(\omega)$ that corresponds to $l \to l' = l \pm 1$ transitions was obtained (see e.g. [27]):

$$\sigma_{kl,l'}^{A@F}(\omega) = |F_{l'}(\omega)|^2 \ \sigma_{kl,l'}^{A}(\omega) \tag{24}$$

where $\sigma_{kl,l'}^{A}(\omega)$ is the pure atomic cross-section for the same transition. The function $|F_{l'}(\omega)|^2$ exhibits minima and maxima, the latter called confinement resonances [40, 27]. No doubt that similar effect will be seen in the cross-section of the fast electron inelastic scattering upon endohedral atom.

For inelastic electron scattering amplitude (6) looks similar to (24):

$$<i | A_L^{A@F}(q,\omega) | f > \cong < i | A_L(q,\omega) | f > F_{l_f}(\omega), \tag{25}$$

In order to take into account the modification of the interaction between the projectile and atom A due to virtual or real excitations of the fulleren $F$ electron shell, one has to solve the equations (5, 8), where $\chi(\omega)$ would include the contribution of the atomic and fulleren excitations on equal grounds. Small size of the atom A as compared to $R$ permits again to separate the atomic and fulleren contribution. Similar to the case of photoionization [28], one has the following expression for the amplitude (6) with account of fulleren effect $A_L^{A@F}(q,\omega)$ expressed via the corresponding pure atomic amplitude:

$$<i|A_L^{A@F}(q,\omega)f> \cong <i|A_L(q,\omega)|f> \left(1 - \frac{\alpha_L^F(\omega,q)}{R^3}\right) \equiv <i|A_L(q,\omega)|f> G_L(\omega,q). \quad (26)$$

Here $\alpha_L^F(\omega,q)$ is the L-pole generalized polarizability of the fullerene shell.

For the case of photoionization the situation is much simpler since only one term - with $L=1$ and in the limit $q=0$, has to be taken into account. Note that $\alpha_1^F(\omega,0) \equiv \alpha_d^F(\omega)$ is the dynamic dipole polarizability of the fullerene. Fortunately, one can determine $\alpha_d^F(\omega)$ almost directly from experiment: $\mathrm{Re}\,\alpha_d^F(\omega) = c\int_I^\infty d\omega' \sigma^F(\omega)/(\omega'^2 - \omega^2)$ and $\mathrm{Im}\,\alpha_d^F(\omega) = c\sigma^F(\omega)/4\pi\omega$, while $\sigma^F(\omega)$ is the fulleren measured photoionization cross-section [41]. However, we cannot do the same for monopole, quadrupole and other multipolarity, as well as for generalized $q$ dependent polarizabilities.

Using (25) and (26) one has the following relation

$$<i|A_L^{A@F}(q,\omega)|f> \cong <i|A_L(q,\omega)|f> F_{l_f}(\omega) G_L(\omega,q). \quad (27)$$

To have a feeling of the impressive role of fulleren shell using as an example the case of $F \equiv C_{60}$, we present $F_{1,2}(\omega)$ in Fig. 7 [42] and $G_1(\omega)$ in Fig. 8 [43].

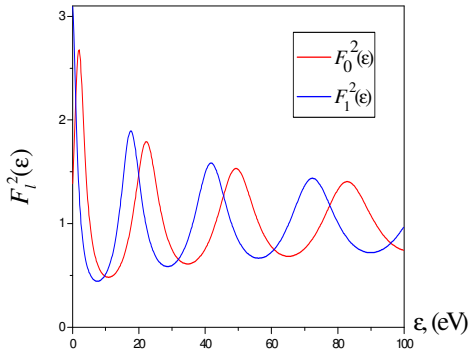

Fig. 7. Factors $F_l(\omega)$ as function of Photoelectron energy $\varepsilon$.

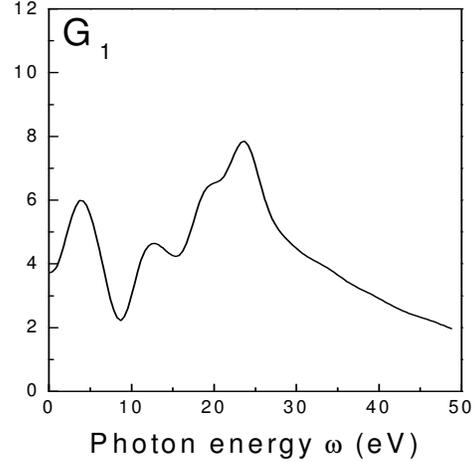

Fig. 8. Dynamic screening factor $G_1(\omega)$.

We see that both factors impressively affect the cross-section. It is reasonable to assume that the corresponding non-dipole factors $G_L(\omega,q)$ will be also considerable and their $q$ dependence, as well as that of $G_1(\omega,q)$ will be essential.

### 9. Conclusions and Perspectives

We have presented here a number of novel results obtained in the investigation on fast electron inelastic scattering and its main characteristics - the GOS. We demonstrated that the GOS are strongly affected by electron correlations. We have discussed intra-doublet correlations, identification of discrete levels, non-dipole corrections to small angle fast electron inelastic scattering, cooperative "drag currents", and Compton scattering. We argued that endohedral atoms are promising new objects for GOS studies.

A number of effects and processes were not discussed mainly due to lack of space. For instance, we did not present reliable arguments (although deserving experimental verification) that the role of correlations in at least dipole GOS is big enough even at high transferred energies. We also did not mention the fact that GOS determine not only electron ionization but also photon emission in the fast electron-atom scattering that is called atomic bremsstrahlung.

Of great interest are such relatively new objects of GOS studies as clusters and fullerenes themselves.

To summarize, let me say that it is a lot to do in this old and still young domain of research.

### 10. Acknowledgement

I am grateful to my co-workers Professors A. S. Baltenkov and L. V. Chernysheva, together with whom some results presented above, were obtained. I acknowledge the financial assistance of the Israeli Science Foundation, grant 174/03 and the Hebrew University Intramural Funds.